\documentclass{ab}

\usepackage{graphicx}
\usepackage{natbib}

\def\pak{$PA_{\rm kin}$\,}
\def\degr{\hbox{$^\circ$}}

\begin{document}

\title{Ionized gas rotation curves in nearby dwarf galaxies}

\author{A.~V.~Moiseev}

\institute{Special Astrophysical Observatory, \frame{Russian Academy of Sciences}\thanks{The system of Russian Academy of Sciences institutes was liquidated (``re-organized'') on Sep 2013},
Nizhnij Arkhyz, 369167, Russia}

\titlerunning{Rotation curves in nearby dwarf galaxies}
\authorrunning{Moiseev}%

\date{September 10, 2013/Revised October 4, 2013}

\offprints{A. Moiseev \email{moisav@sao.ru}}

\abstract{
We present the results of study of the ionized gas velocity fields
in 28 nearby (systemic velocity below $1000$~km\,s$^{-1}$) dwarf
galaxies.  The observations were made at the 6-m BTA telescope of
the SAO~RAS with the scanning Fabry--Perot interferometer in the
H$\alpha$ emission line. We were able to measure regular circular
rotation parameters  in 25~galaxies. As a rule, rotation
velocities measured in H\,II are in a good agreement with the data
on the  H\,I  kinematics  at the same radii. Three galaxies reveal
position angles of the kinematic axis in the  H\,II velocity
fields that strongly (tens of degrees) differ from the
measurements in neutral hydrogen at large distances from the
center or from the orientation of the major axis of optical
isophotes. The planes of the gaseous and stellar disks in these
galaxies most likely do not coincide. Namely, in DDO\,99 the
gaseous disk is warped beyond the optical radius, and in UGC\,3672
and UGC\,8508 the inclination of orbits of gas clouds varies in
the inner regions of galaxies. It is possible that the entire
ionized gas in UGC\,8508 rotates in the plane polar to the stellar
disk.
}

\maketitle

\section{INTRODUCTION}
\label{intro}

Rotation curve is one of the main characteristics of galaxies,
accessible to direct measurements. Knowing the radial distribution
of rotation velocity, we can, given certain approximations,
estimate the  distribution of mass, including that of dark matter.
It is rather difficult to measure the rotation curve of gas in
dwarf galaxies. This is due to their small mass, and hence the
velocity of regular rotation is only a few tens of~km\,s$^{-1}$
which is comparable to the amplitude of noncircular gas motions
related to the regions of current star formation. It is often not
enough here to only have to cross-section the line of sight velocity
distribution along the major axis of the disk, but we
additionally need to analyze the two-dimensional distribution of
line of sight velocities (the velocity field) to be able to fit it using
the  model of circular rotation.

Therefore, most of the  reliable rotation curves of gas in dwarf
galaxies published are obtained from the H\,I observations in the
21~cm line, applying the aperture synthesis method. However, even
using sufficiently large databases, the angular resolution of
these observations is not very large, e.g., amounting to
$13$--$61''$ in the  FIGGS survey~\citep{Begum2008}.
Recent observations for the LITTLE~THINGS
sample~\citep{Hunter2012} were obtained with a resolution
of $6$--$10''$, which can be considered record high in this kind
of studies. However,  many scientific problems require the
knowledge of gas kinematics with a better spatial resolution. For
instance, this is the case with the study of the central regions
of dwarf galaxies, where the divergence of the actual rotation
curve from the predictions of numerical calculations is the most
conspicuous \citep[the nuclear cusp problem, see, e.g., the review by][] {Doroshkevich2012UFN}.

The observations with the scanning Fabry--Perot interferometer
(FPI) in the ionized gas emission lines allow to construct radial
velocity fields with the seeing-limited  angular resolution
(usually $1$--$2''$). However, the disadvantage of the H\,II line
measurements, as compared with H\,I, is a smaller filling of the
velocity field, since the emission in the Balmer lines is visible
only where there is a sufficient number of UV photons, the main
source of which are young, massive stars. In order to detect
diffuse gas emission away from the regions of star formation,
large optical telescopes or hours of scanning cycles at the
instruments of small diameters are required. Therefore, most of
the observations of dwarf galaxies with the IFP were carried out
for the objects with a relatively strong star formation, such as
the blue compact
galaxies~\citep{Ostlin1999,Lozinsk2006}. About
two dozen dIrr galaxies were  also observed within the GHASP
survey~\citep{Epinat2008}, but an insufficient limit of
these observations has often not allowed to build the rotation
curves in the galaxies deprived of bright  H\,II regions.

In this regard, combining the rotation curves obtained for the
central regions of galaxies from the ionized gas data with the
results from neutral hydrogen data for the more external regions
appears to be promising. This paper is one of the steps in this
direction. We present here the reduction results of the archival
observations of nearby dwarf galaxies (systemic velocity of less
than $1000$~km\,s$^{-1}$), performed with a scanning FPI on the
6-m BTA telescope of the Special Astrophysical Observatory of the
Russian Academy of Sciences. Most of the studied objects were
observed within the proposal of A.~A.~Klypin, and the rest of
them---on the requests of A.~Begum, T.~A.~Lozinskaya, and
S.~A.~Pustilnik. Further, Section~\ref{sec_obs}
describes the observations and data reduction,
Section~\ref{sec_anal} discusses the technique of
construction of rotation curves, Section~\ref{sec_res}
presents our results (velocity fields and rotation curves), and
finally, Section~\ref{sec_notes} gives comments on the
kinematic features of individual galaxies.

\section{OBSERVATIONS AND DATA REDUCTION }
\label{sec_obs}

The observations were conducted in the primary focus of the
\mbox{6-m} BTA telescope of the SAO RAS with the scanning FPI
installed inside the SCORPIO focal
reducer~\citep{AfanasievMoiseev2005}. The operating
spectral range around the H$\alpha$ line was cut by the
narrow-band filters with a bandwidth of \mbox{${\rm
FWHM}=14$--$21$~\AA}. Until November 2009 the observations were
made with the  IFP\,501 interferometer providing in the H$\alpha$
line a free spectral range between the adjacent interference
orders of  $\Delta\lambda=13$~\AA\, and spectral resolution ($\rm
FWHM$ of the instrumental profile) of about $0.8$~\AA\,
($35$~km\,s$^{-1}$) at the scale of  0.36~\AA\, per channel. Later
we used the  IFP\,751 interferometer having
$\Delta\lambda=8.7$~\AA\, and spectral resolution of $0.4$~\AA\,
($18$~km\,s$^{-1}$) at the scale of  0.21~\AA\, per channel.

The   detectors used in  2005--2010  were the \linebreak
EEV\,42-40 and E2V\,42-90 CCDs, providing the image scale of
$0\farcs71$~pixel$^{-1}$ in the $4\times4$   binning readout mode.
In 2002 the  TK\,1024 CCD was used as a detector, providing the
image scale of $0\farcs56$~pixel$^{-1}$ in the $2\times2$ binning
mode.

In the scanning process we have consistently obtained the object
interferograms (36 for IFP\,501 and 40 for IFP\,751) at different
distances between the  IFP plates. Reduction of the observational
material was performed using the software package operating in the
IDL environment~\mbox{\citep{Moiseev2002ifp,MoiseevEgorov2008}}. The reduction result is a data cube
where each pixel in the field of view contains a 36- or 40-channel
spectrum.

The log of observations is given in Table~\ref{tab_obs}
containing: the name of the galaxy (adopted in the paper and the
alternative, if it is commonly used), the date of observations,
the interferometer name, exposure time, image quality (seeing),
and final angular resolution ($\omega$) after smoothing the
reduced data cubes by a two-dimensional Gaussian  to increase the
signal-to-noise ratio in the regions of low surface brightness.

\begin{table*}[tbp]
\captionstyle{normal} \caption{Log of observations}
\label{tab_obs}
\medskip
\begin{tabular}{l|l|r|c|c|c|c}\hline
\multicolumn{1}{c|}{Name} &  Alt. name   & \multicolumn{1}{c|}{Date}    & FPI       & Exp. time, sec   & Seeing, arcsec &    $\omega$, arcsec\\
\hline
CGCG\,269-049&                & Feb 06, 2010 & IFP\,751 & $150\times40$ &  3.5--4.4 &4.4  \\
DDO\,53      &                & Feb 26, 2009 & IFP\,501 & $200\times36$ &  2.3--2.5 &3.3   \\
DDO\,68      &                & Dec 30, 2006 & IFP\,501 & $240\times36$ &  1.5--2.7 &2.7   \\
DDO\,99      &                & Feb 26, 2009 & IFP\,501 & $180\times36$ &  2.0--3.0 &3.8   \\
DDO\,125     &                & May 18, 2005 & IFP\,501 & $180\times36$ &  1.5--1.8 &3.0   \\
DDO\,190     &                & Mar 04, 2009 & IFP\,501 & $100\times36$ &  2.2--2.5 & 3.3   \\
KK\,149      &  [KK\,98] 149  & Mar 05, 2009 & IFP\,501 & $150\times36$ &  1.9--2.5 & 2.8   \\ 
KKH\,12      &                & Aug 23, 2004 & IFP\,501 & $120\times36$ &  1.8--2.0 & 2.7   \\
KKH\,34      &                & Nov 12, 2009 & IFP\,751 & $230\times40$ &  1.9--3.1 & 3.4   \\
KKR\,56      &                & May 20, 2010 & IFP\,751 & $150\times40$ &  2.5--3.0 & 3.3   \\
UGC\,231     &  NGC\,100      & Nov 11, 2009 & IFP\,751 & $200\times40$ &  1.4--2.3 & 2.4   \\
UGC\,891     &                & Nov 11, 2009 & IFP\,751 & $200\times40$ &  2.0--3.1 & 3.1   \\
UGC\,1281    &                & Aug 14, 2009 & IFP\,751 & $110\times40$ &  1.5--1.7 & 2.7   \\
UGC\,1501    &  NGC\,784      & Nov 10, 2009 & IFP\,751 & $200\times40$ &  1.4--2.0 & 2.3   \\
UGC\,1924    &                & Nov 11, 2009 & IFP\,751 & $180\times40$ &  1.4--2.0 & 2.3   \\
UGC\,3476    &                & Nov 02, 2010 & IFP\,751 & $220\times40$ &  1.4--1.8 & 2.4   \\
UGC\,3672    &                & Nov 12, 2009 & IFP\,751 & $160\times40$ &  1.7--3.1 & 3.1   \\
UGC\,5423    &  M\,81\,dwB    & Feb 26, 2009 & IFP\,501 & $180\times36$ &  2.7--2.9 & 3.5   \\
UGC\,5427    &                & Mar 04, 2009 & IFP\,501 & $180\times36$ &  2.2--2.8 & 3.7   \\
UGC\,6456    & V\,II\,Zw\,403 & Nov 29, 2002 & IFP\,501 & $300\times36$ &  1.5--2.1 &  2.2   \\
UGC\,7611    &  NGC\,4460     & May 19, 2010 & IFP\,751 & $160\times40$ &  2.1--2.5 & 3.5    \\
UGC\,8508    &                & May 16, 2005 & IFP\,501 & $200\times36$ &  1.5--2.0 &  3.0   \\
UGC\,8638    &                & Feb 24, 2009 & IFP\,501 & $150\times36$ &  2.8--3.5 & 3.9   \\
UGC\,11425   &  NGC\,6789     & Aug 14, 2009 & IFP\,751 & $140\times40$ &  1.5      & 3.0   \\
UGC\,11583   &                & Nov 10, 2009 & IFP\,751 & $220\times40$ &  1.2--1.6 & 1.9   \\
UGC\,12713   &                & May 16, 2005 & IFP\,501 & $200\times36$ &  3.1--3.5 &  3.0   \\
UGCA\,92     &                & Nov 10, 2009 & FPI\,751 & $180\times40$ &  1.4--1.9 &  2.5   \\
UGCA\,292    &  CVn\,I\,dwA   & Feb 07, 2010 & IFP\,751 & $120\times40$ &  2.4--3.1 & 3.6   \\
\hline
\end{tabular}
\end{table*}

The observed H$\alpha$ line profiles were approximated by the
Voigt function, which describes them well enough  in most of
cases. According to the results of profile approximations, we have
built the ionized gas line of sight velocity fields and line of sight velocity
dispersion maps as well as the images of galaxies in the H$\alpha$
emission line and in the continuum. Some of these maps for seven
galaxies of our list have already been presented in our previous
paper~\citep{MoisLoz2012}, where we discuss the features
of distribution of the  ionized gas velocity dispersion.

\section{KINEMATICS ANALYSIS}
\label{sec_anal}

\subsection{Moderately Inclined Disks}

The inclination of the disk plane to the line of sight $i_0$ does
not exceed $70$\degr{} for most of the galaxies considered. We can
neglect the thickness of the gaseous disk here and approximate the
observed velocity fields within the commonly used tilted-rings
approximation. We used the software package, written in the IDL,
implementing the adaptation of this method for the  ionized gas
velocity fields \citep[][ibid references to earlier studies]{Moiseev2004,Moiseev2008}).

During the analysis,  the observed velocity field is split into
$1$--$5''$-wide elliptical rings  in agreement with the adopted
inclination $i_0$ and  position angle of the major axis of the
disk ${\rm PA}_0$. In each ring, we fit the   observed
distribution of line of sight velocities by  the  circular rotation
model, using the $\chi^2$-minimization. Model parameters are the
position angle of the kinematic major axis \pak, the inclination
of circular orbits $i$, circular rotation velocity $V_{\rm rot}$,
and systemic velocity $V_{\rm sys}$. If one can be sure that the
disk plane has no large warps, we can accept the inclination and
systemic velocity to be radius-independent (\mbox{$i(r)=i_0$},
\mbox{$V_{\rm sys}(r)={\rm const}$}). In this case, radial
variations of \pak\ reflect the behavior of non-circular
components of the three-dimensional velocity vector of gas clouds,
for example, as a result of disturbance by the spiral density
waves.

Expanding shells, induced by star formation in dwarf galaxies, can
significantly distort the velocity field of the disk, since the
amplitude of the rotation curve is small and comparable to the
shell expansion velocity \mbox{($10$--$30$~km\,s$^{-1}$)}, while
the size of the shell may be comparable to the size of the disk.
We have shown in~\cite{Moiseev2010} that in this case it
is necessary to fix the  $V_{\rm sys}$, $i$ and \pak\ parameters
to avoid systematic errors in the measurement of rotation
velocities by the tilted-rings method.

This is why the analysis of the velocity fields is carried out as
follows. As the first approximation for ${\rm PA}_0$, $i_0$ and
the center of rotation, we have adopted the existing orientation
parameters according to the photometry data\footnote{If the
authors listed the ellipticity isophotes only, the corresponding
$i_0$ were calculated according to the
relation~\cite{Staveley-Smith1992} for dwarf galaxies.}
or H\,I maps. Their further refinement is done by successive
approximations, while at each step we masked the regions notable
for their peculiar kinematics, i.e., deviating by more than
\linebreak $6$--$10$~km\,s$^{-1}$  from the model of circular
rotation (this threshold was individually selected). If the
center of symmetry of a   velocity field lies no further
than  $1$--$3''$  from the galactic core (the center of internal
isophotes), then the  photometric center was fixed as the
center of circular rotation. For $V_{\rm sys}$ and ${\rm PA}_0$
we took the mean values of these parameters along the radius. In
some cases, where we suspect a warp of the gas disk, variations
of \pak\ with radius were allowed.

The inclination of the disk to the line of sight was usually fixed
based on the data on the morphology of the galaxy, since the $i_0$
estimates from the velocity fields with low rotation amplitudes are  uncertain. In some cases, we still managed to estimate $i_0$
within the assumption  of the approach of pure circular rotation.
In general, we tried to maximize the agreement in the orientation
parameters determined according to various sources: optical
imagery, H\,I maps, velocity fields of ionized gas (see the
remarks in Section~\ref{sec_notes}).

\subsection{Disks observed edge-on}

The sample includes five galaxies  (UGC\,231, UGC\,1281,
UGC\,1924, UGC\,3476,  UGC\,11583), with disk inclination almost
exactly corresponding to the edge-on orientation, i.e.
$i_0>88\degr$, according the  available  images. The thin disk
model is obviously not suitable for the description of their
velocity fields. In the general case, we have to construct a 3D
model of the rotating disk, taking into account the effects
produced on the observed velocity field by its vertical
structure, internal dust absorption, etc., as done, for example,
by \cite{Kamphuis2007}. However,
construction of such complex models requires additional
information on the morphology of galaxies at different
wavelengths, which is not always available. This is why we used a
simple approximation of a transparent cylinder rotating around
the axis, wherein the observed  line of sight velocity $V_{\rm obs}$ is
determined, mostly, by the emission of regions located on the
line passing through the center of the galaxy perpendicular to
the line of sight. So that:
\begin{equation}
\begin{array}{lcl}
 V_{\rm obs}(x,z) &=&V_{\rm rot}(r)+V_{\rm sys}, \\
 V_{\rm rot}(-r) &=&-V_{\rm rot}(r), \quad r=x.
\label{eq1}
\end{array}
\end{equation}

The $x$ axis here is directed from the center of the galaxy along
the major axis of the disk, while $z$ is oriented orthogonally to
it. This means that the shape of the rotation curve is preserved
with the distance from the disk plane. We consider this
approximation valid, since the observed emission disks of the
studied galaxies are relatively thin (except perhaps UGC\,3476).
Internal dust absorption in the studied dwarf galaxies is low. At
least, no contrasting dust lanes are present. Another
simplification of the real situation is that
ratio~(\ref{eq1}) does not account for the contribution
of all emitting regions located in the line of sight. \citet{ZasovKhoperskov2003} show that this
effect is important in the central regions of galaxies, since it
makes the observed rotation velocity gradient less steep which is
especially critical for the galaxies with the central maximum on
the rotation curve. But for the flat or solid-body rotation
curves, typical of dwarf galaxies, the distortions of the real
picture are large, especially in the outer regions.

The observed velocity fields were described by the two-dimensional
model defined by expression~(\ref{eq1}), the fit
parameters were the position of the rotation center, ${\rm PA}_0$
and $V_{\rm sys}$. The model fitting was done by the same method
as in the case of moderately inclined disks:
$\chi^2$-minimization, masking of regions with peculiar
kinematics, comparison of the positions of the kinematic and
photometric centers, and so on~(see the previous section). When
calculating the rotation curve, the step in $r$ was usually about
$2''$ in the center, reaching $8$--$10''$ in the outer regions.

\section{RESULTS}
\label{sec_res}

Figure~\ref{fig_1} along with the original data shows
the model velocity field and the result of subtracting the model
from the observations. In cases where we did not manage to build a
rotation model, only the initial data without subtracting the
systemic velocity are listed.

\begin{figure*}[tbp!!!]
\centerline{\includegraphics[width=18. cm]{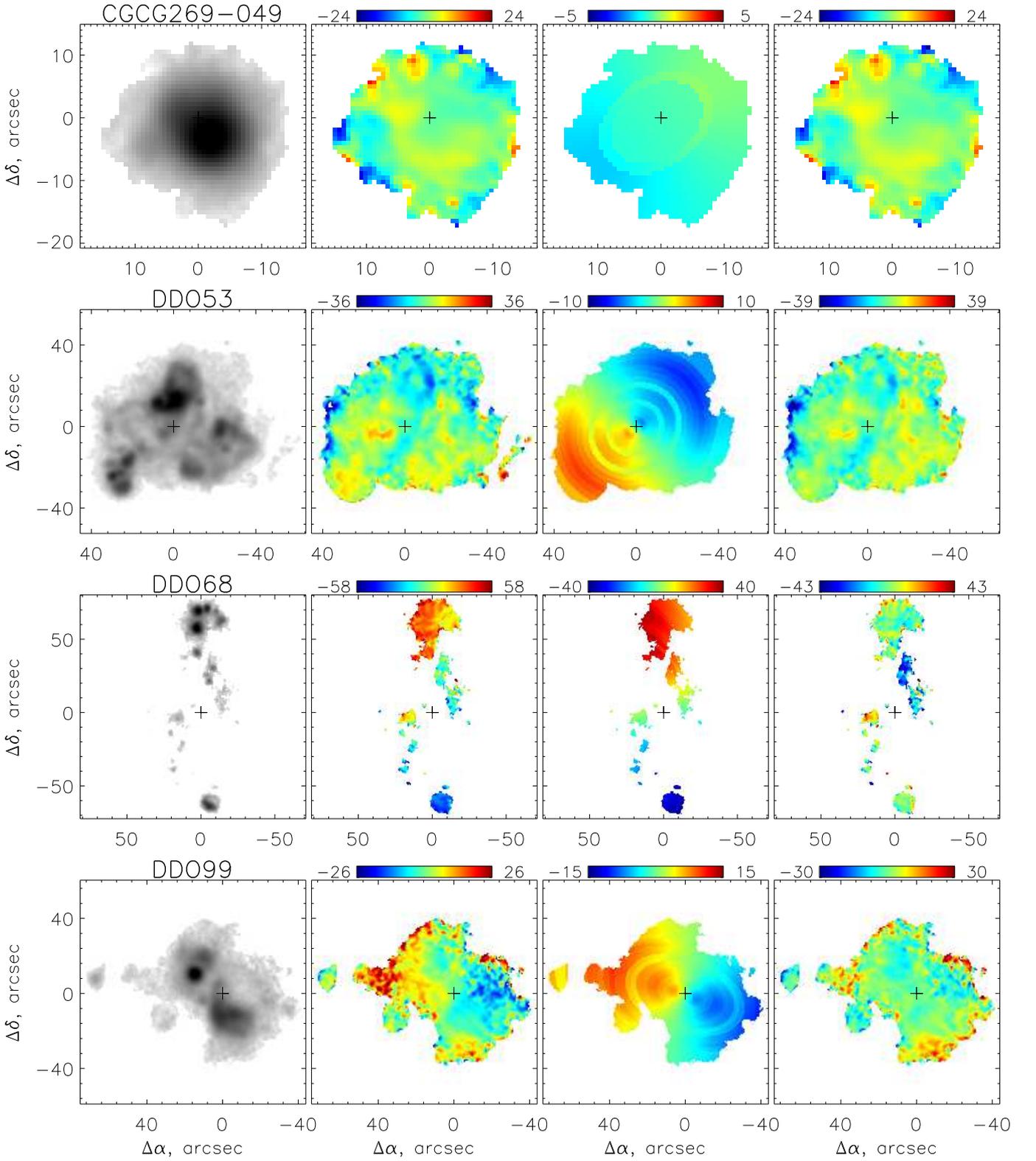}}
\caption{The results of observations and analysis within the
assumption of   circular rotation. From left to right: the
H$\alpha$ line image, the line of sight velocity field (systemic
velocity $V_{\rm sys}$, listed in Table~\ref{tab_res},
is subtracted), model velocity field, residual velocity field
(observations minus model). The cross marks the position of the
kinematic center. The scale is in km\,s$^{-1}$. }
\label{fig_1}
\end{figure*}

\setcounter{figure}{0}

\begin{figure*}[tbp!!!]
\centerline{\includegraphics[width=18. cm]{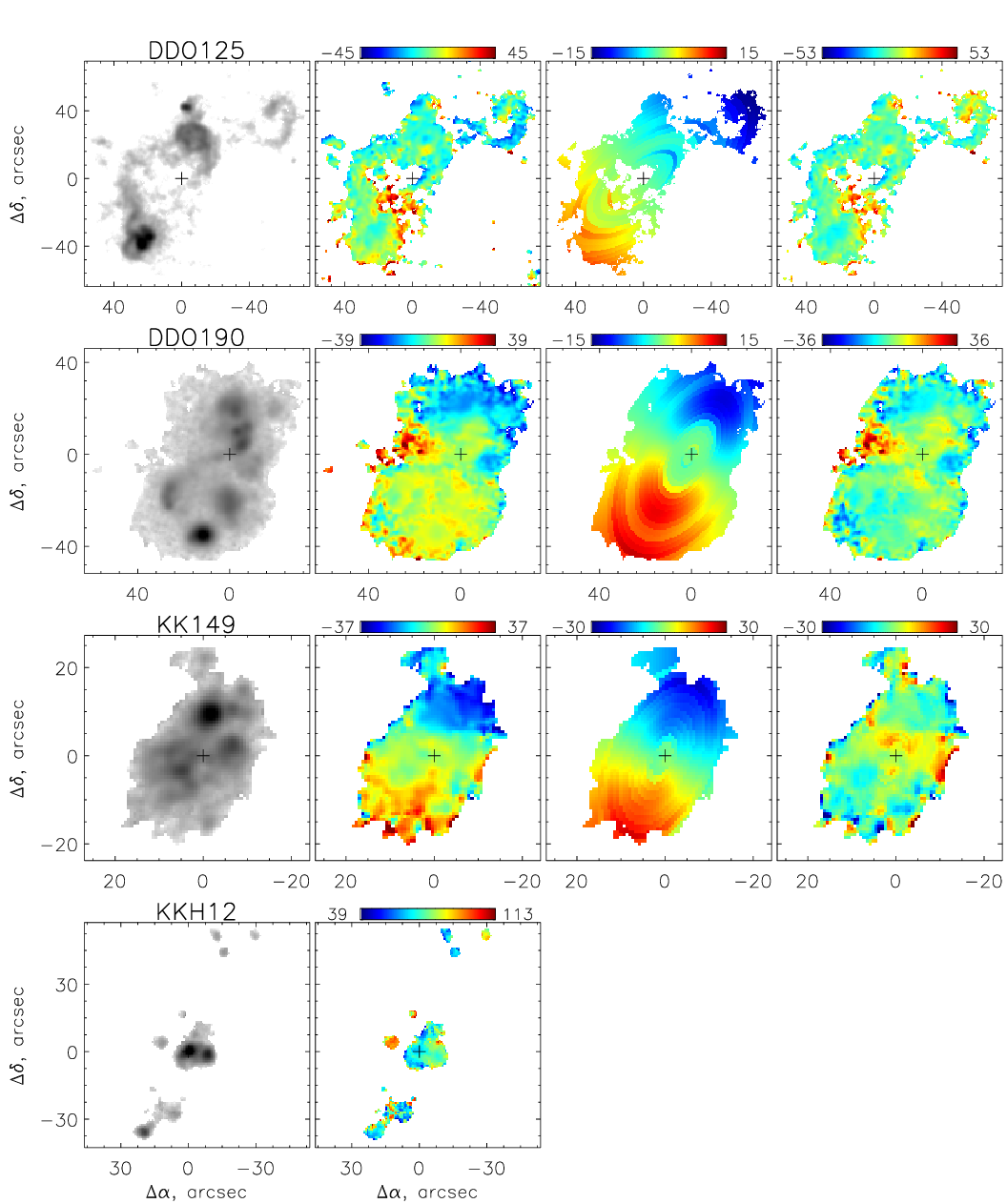}}
\caption{(Contd.)}
\end{figure*}

\setcounter{figure}{0}

\begin{figure*}[tbp!!!]
\centerline{\includegraphics[width=18. cm]{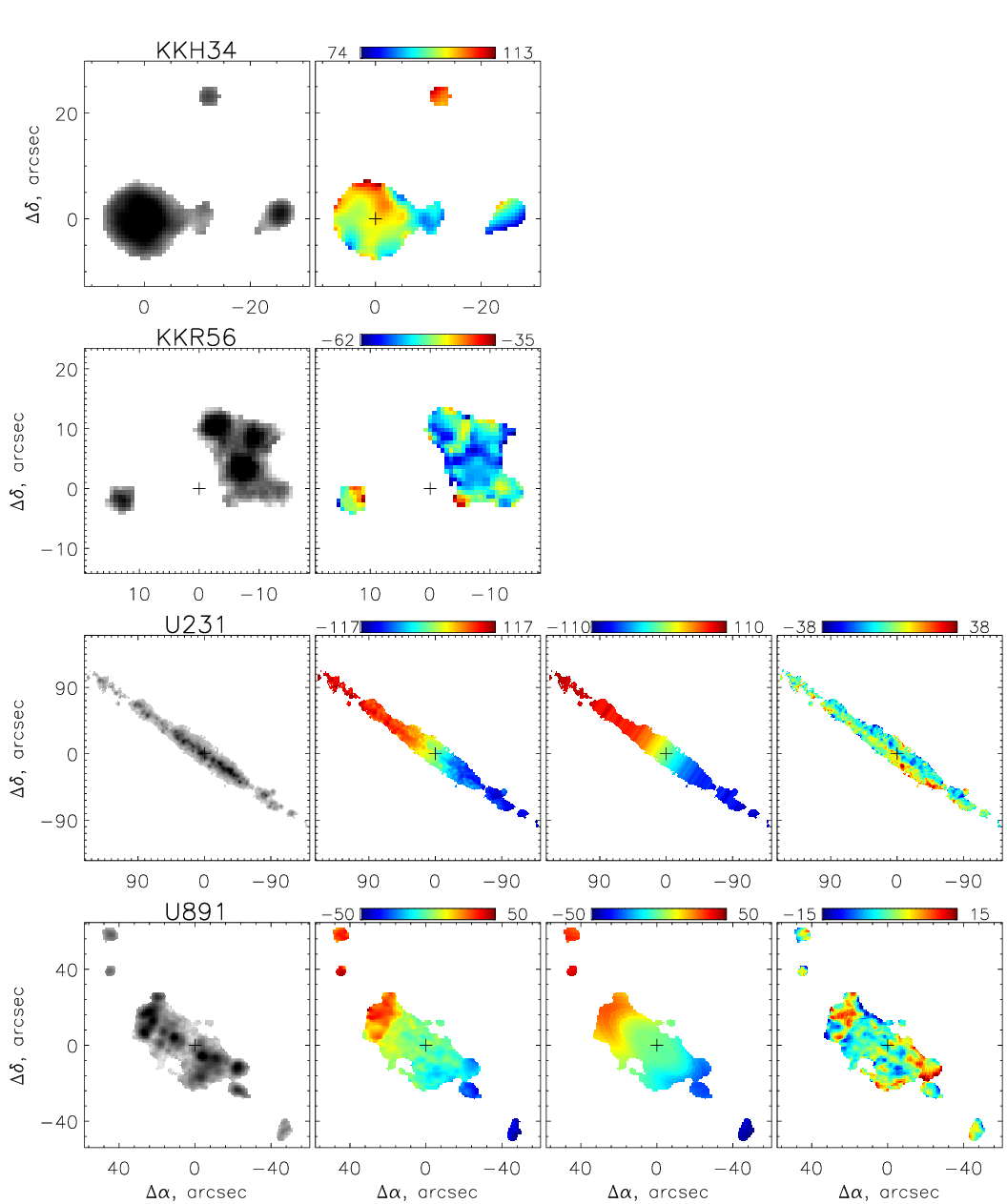}}
\caption{(Contd.)}
\end{figure*}

\setcounter{figure}{0}

\begin{figure*}[tbp!!!]
\centerline{\includegraphics[width=18. cm]{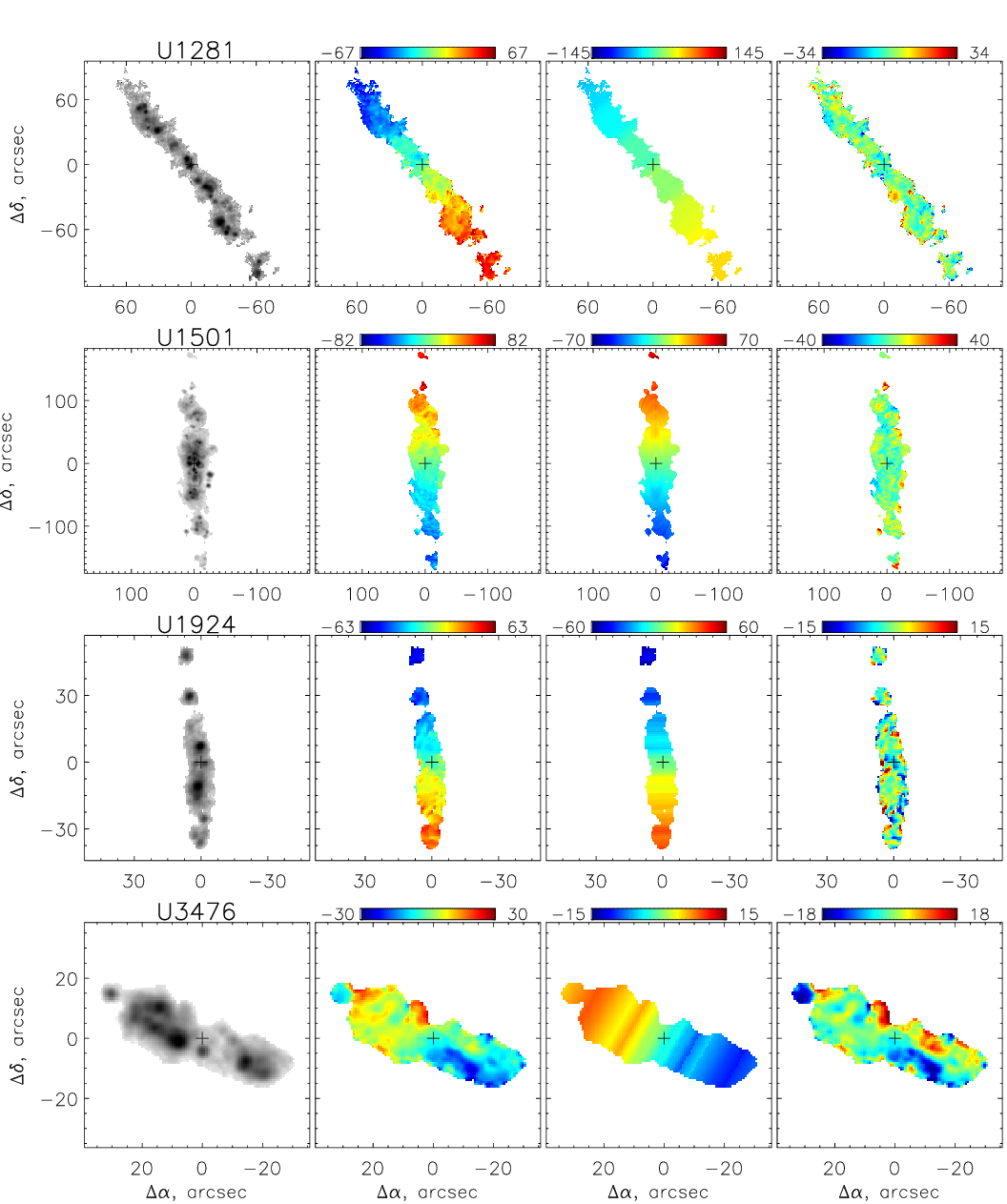}}
\caption{(Contd.)}
\end{figure*}

\setcounter{figure}{0}

\begin{figure*}[tbp!!!]
\centerline{\includegraphics[width=18. cm]{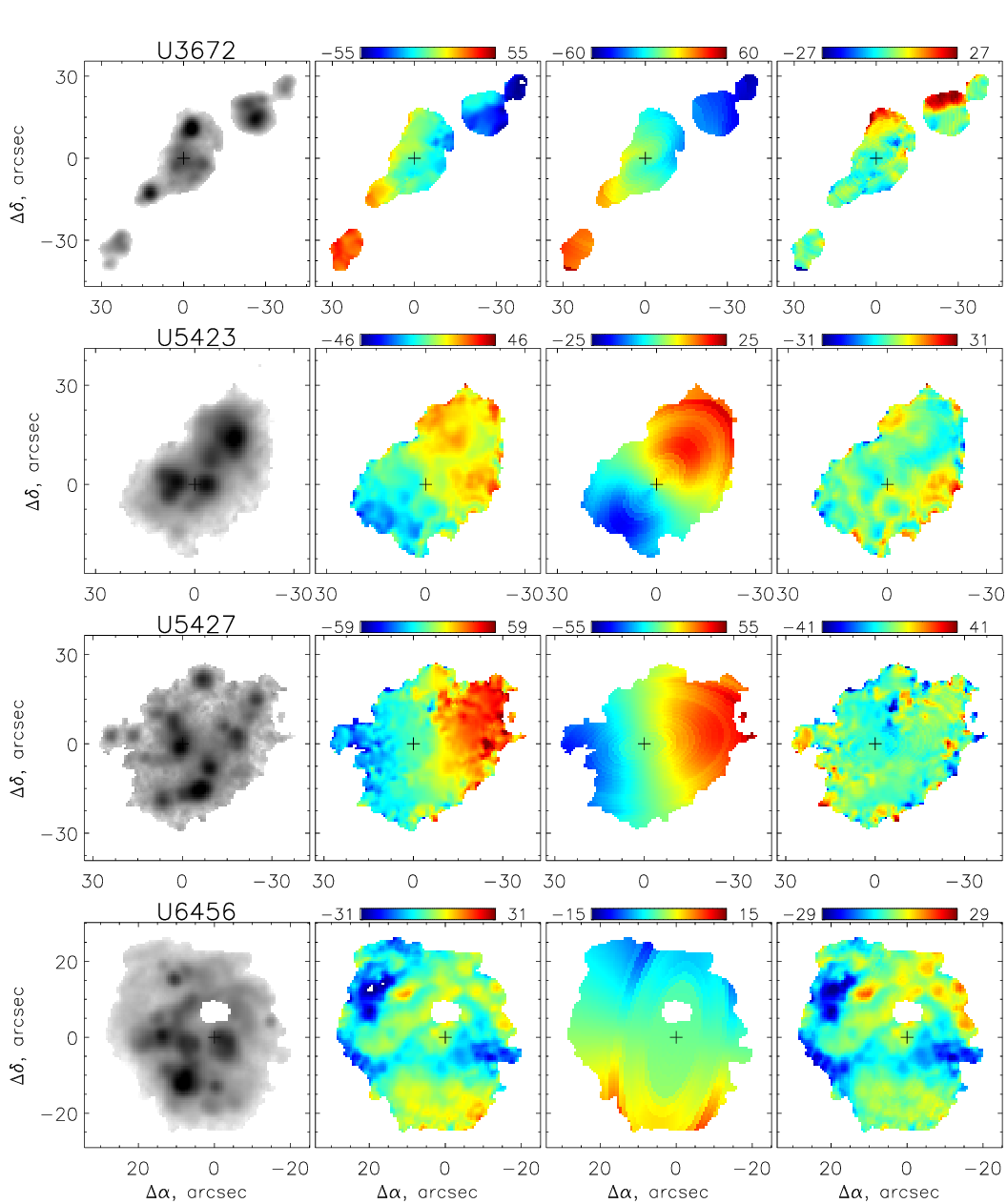}}
\caption{(Contd.)}
\end{figure*}

\setcounter{figure}{0}

\begin{figure*}[tbp!!!]
\centerline{\includegraphics[width=18. cm]{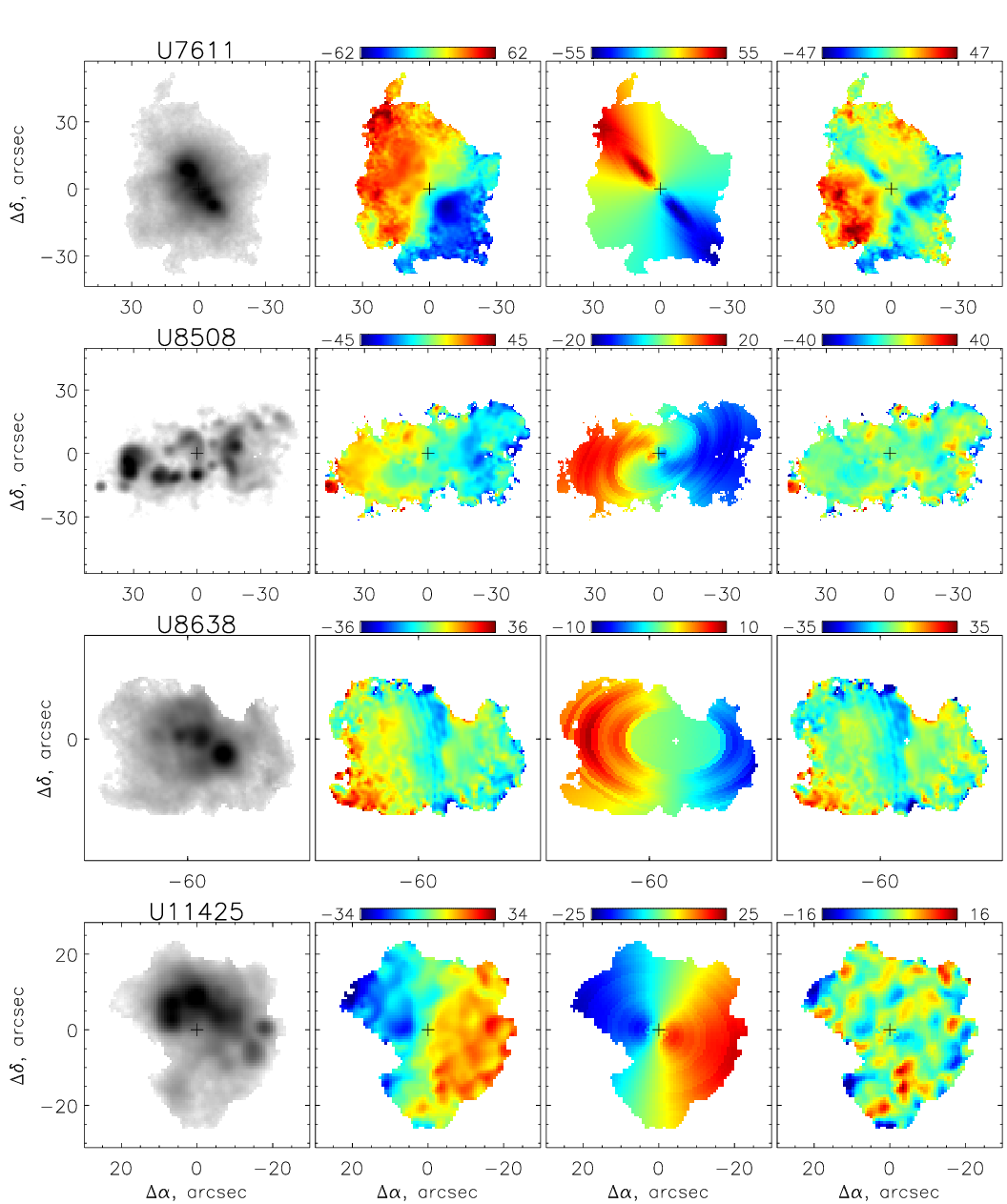}}
\caption{(Contd.)}
\end{figure*}

\setcounter{figure}{0}

\begin{figure*}[tbp!!!]
\centerline{\includegraphics[width=18. cm]{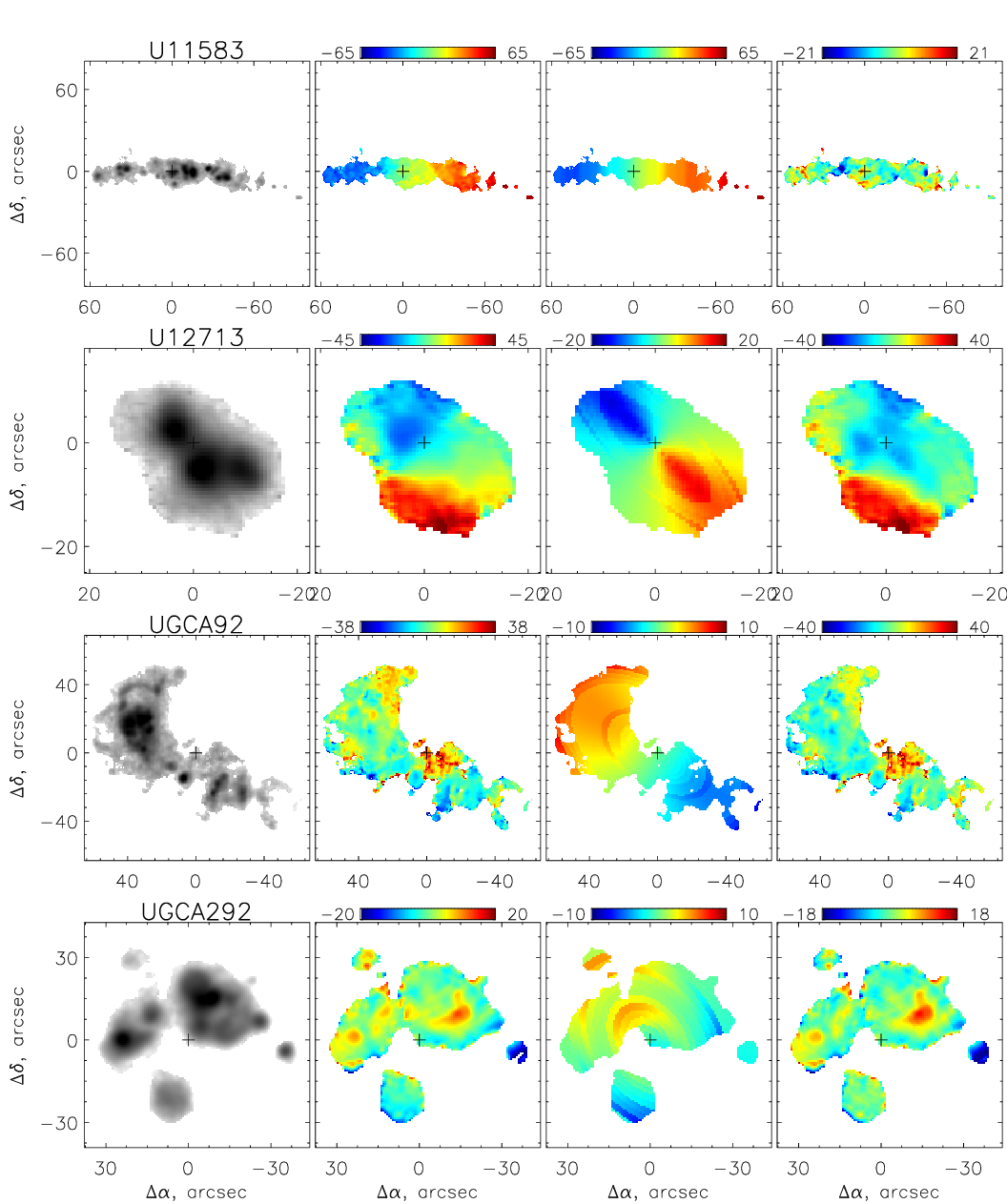}}
\caption{(Contd.)}
\end{figure*}

\begin{figure*}[tbp!!!]
\centerline{\includegraphics[width=5.8 cm]{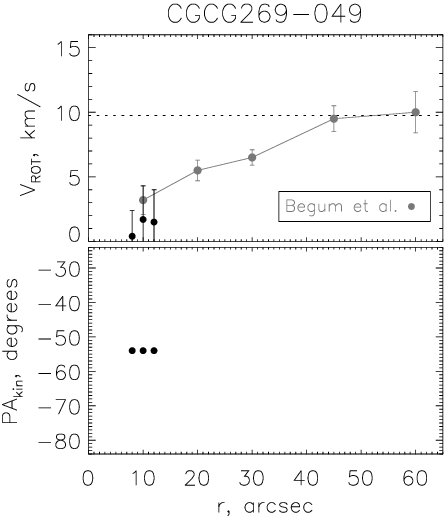}
\includegraphics[width=5.8 cm]{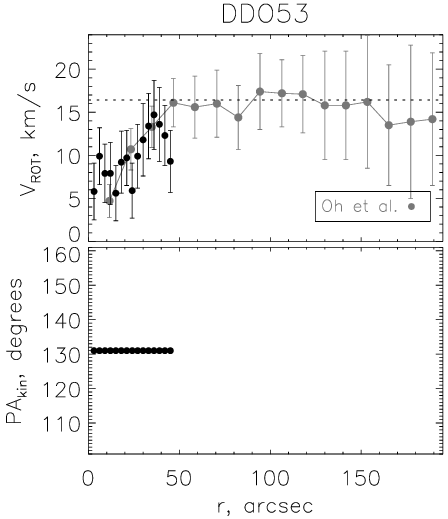}
\includegraphics[width=5.8 cm]{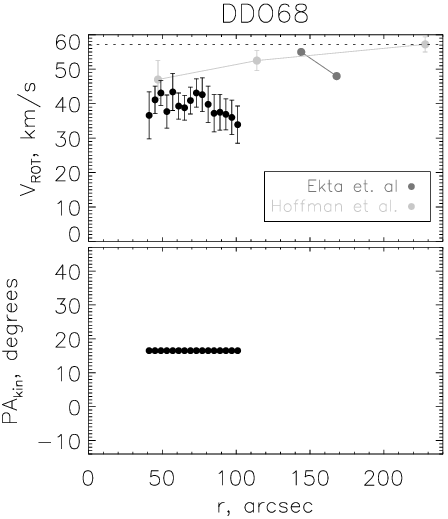}
}
 \vspace{3mm}
\centerline{\includegraphics[width=5.8 cm]{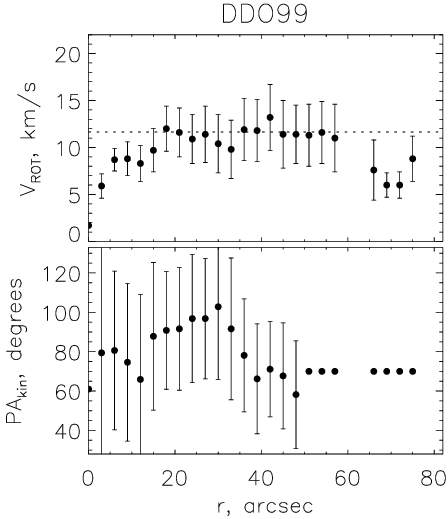}
\includegraphics[width=5.8 cm]{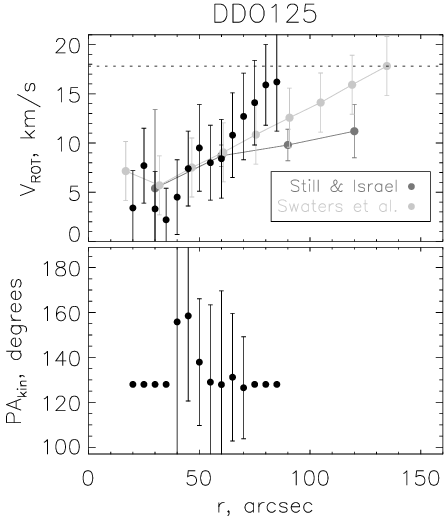}
\includegraphics[width=5.8 cm]{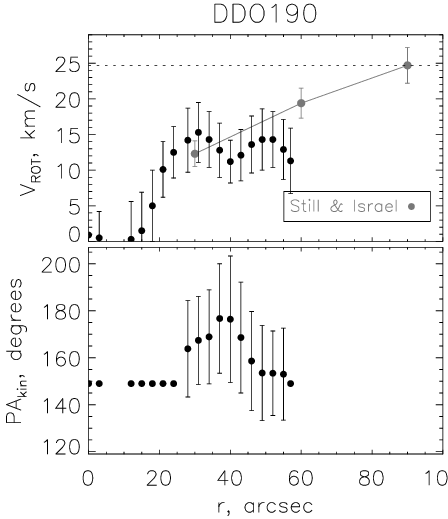}
}
 \vspace{3mm}
\centerline{\includegraphics[width=5.8 cm]{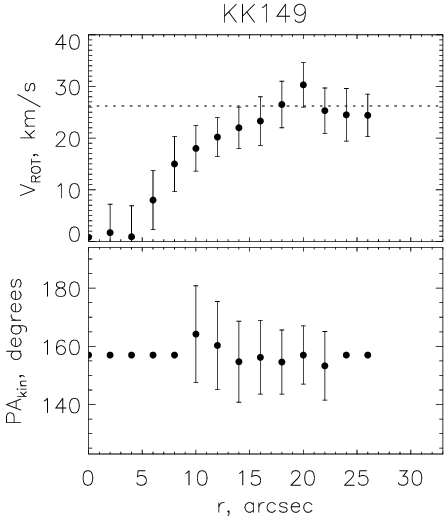}
\includegraphics[width=5.8 cm]{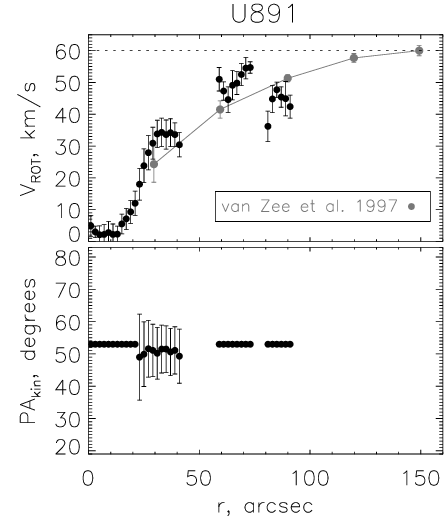}
\includegraphics[width=5.8 cm]{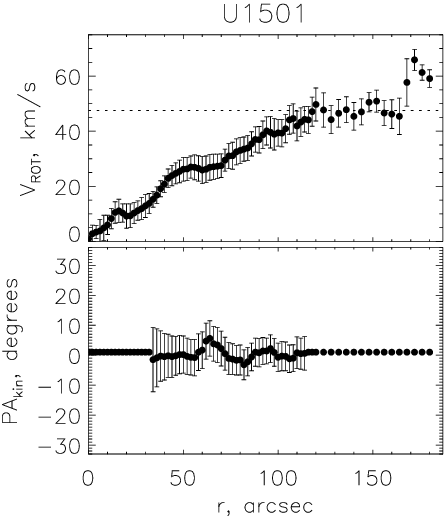}
} \caption{Rotation curves and radial variations of \pak\ for the
galaxies with moderately inclined  disks. Black circles represent
our measurements from the velocity fields in H$\alpha$, gray
circles mark the literature data. The dashed line marks the
adopted $V_{\rm max}$ value.} \label{fig_rc1}
\end{figure*}

\setcounter{figure}{1}

\begin{figure*}[tbp!!!]
\centerline{\includegraphics[width=5.8 cm]{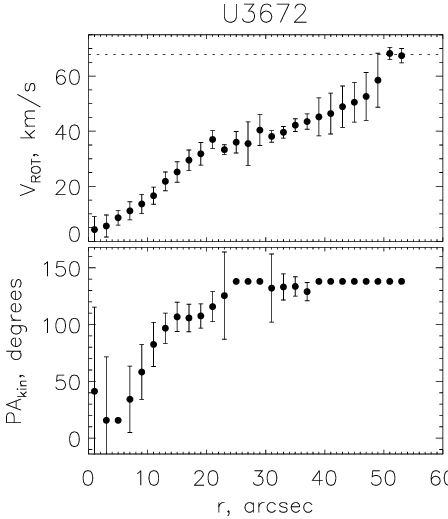}
\includegraphics[width=5.8 cm]{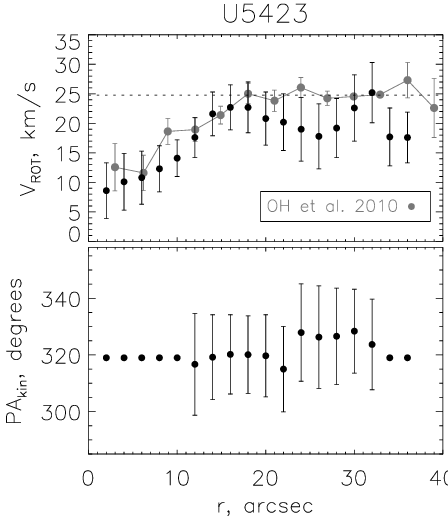}
\includegraphics[width=5.8 cm]{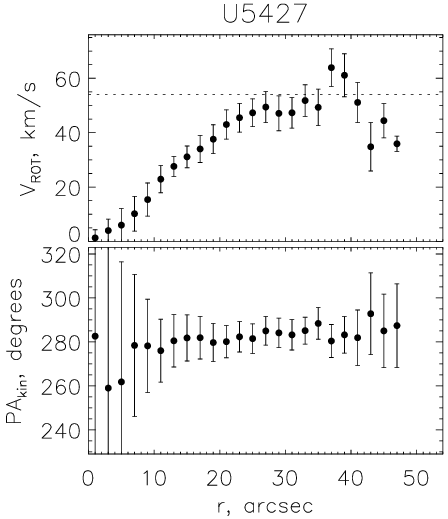}
}
 \vspace{3mm}
\centerline{\includegraphics[width=5.8 cm]{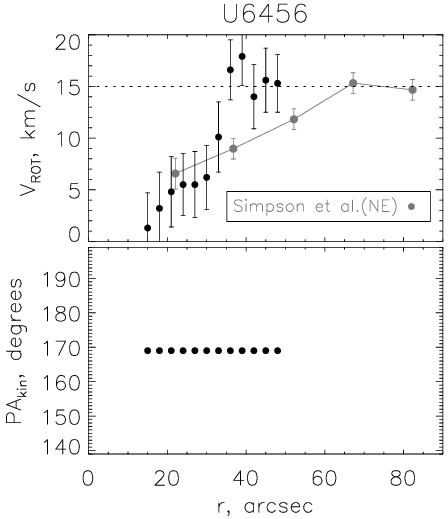}
\includegraphics[width=5.8 cm]{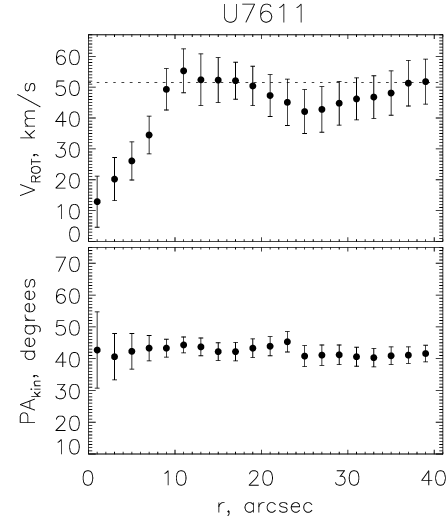}
\includegraphics[width=5.8 cm]{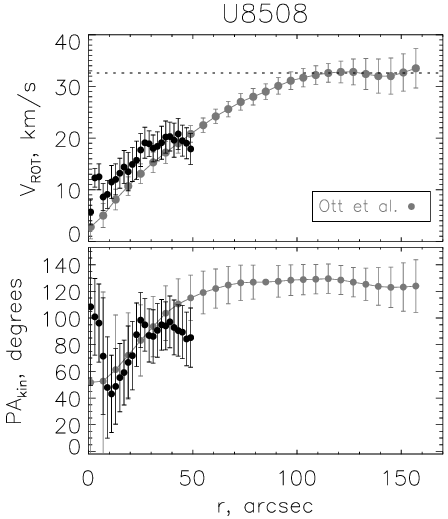}
}
 \vspace{3mm}
\centerline{\includegraphics[width=5.8 cm]{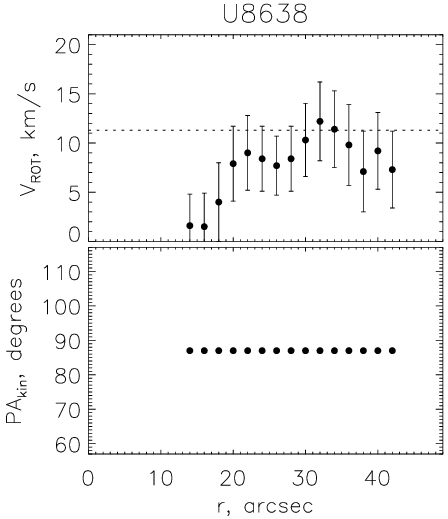}
\includegraphics[width=5.8 cm]{Moiseev_fig2_16.eps}
\includegraphics[width=5.8 cm]{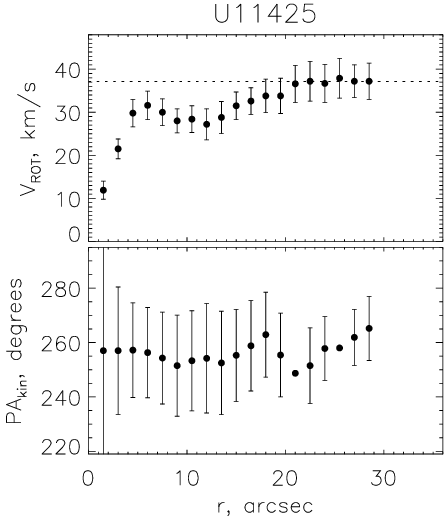}
} \caption{ (Contd.)}
\end{figure*}

\setcounter{figure}{1}

\begin{figure*}[tbp!!!]
\centerline{\includegraphics[width=5.8 cm]{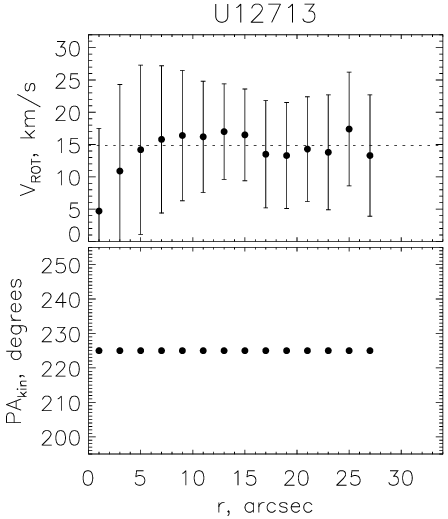}
\includegraphics[width=5.8 cm]{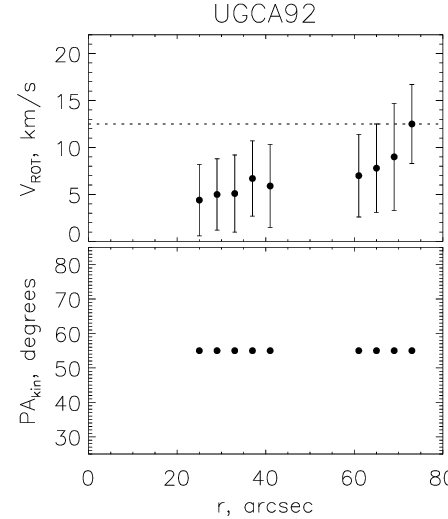}
\includegraphics[width=5.8 cm]{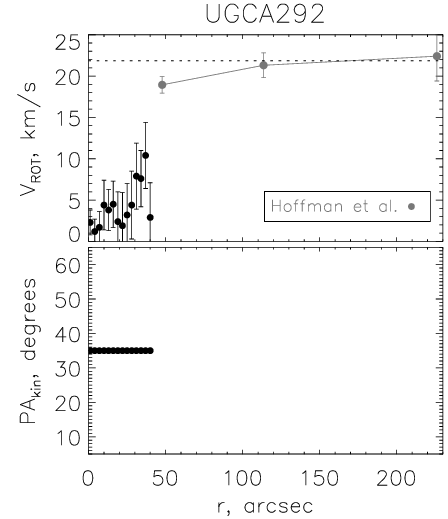}
} \caption{(Contd.)}
\end{figure*}

\begin{figure*}[tbp!!!]
\includegraphics[width=5.8 cm]{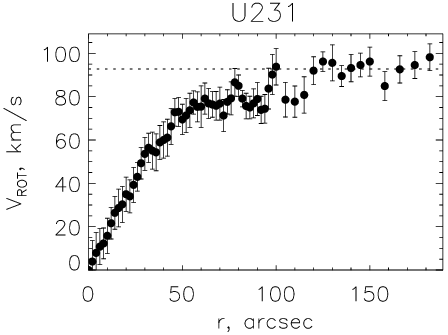}
\includegraphics[width=5.8 cm]{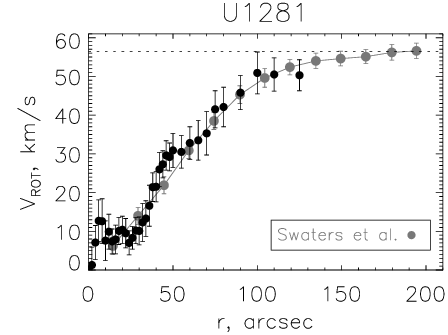}
\includegraphics[width=5.8 cm]{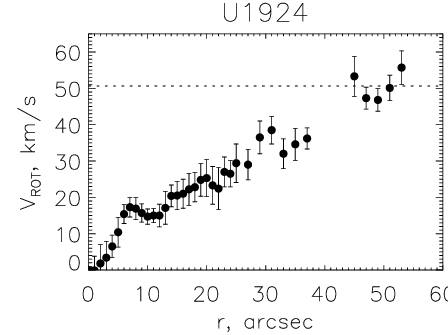}\\
\includegraphics[width=5.8 cm]{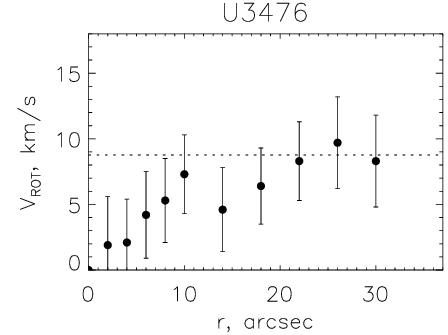}
\includegraphics[width=5.8 cm]{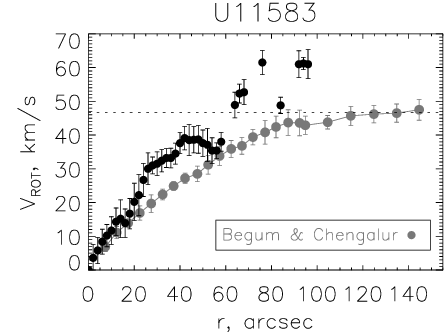}
\caption{Rotation curves of the edge-on galaxies. The markings are
the same as in Fig.~\ref{fig_rc1}.}
\label{fig_rc2}
\end{figure*}

Figure~\ref{fig_rc1} shows the rotation curves $V_{\rm
rot}(r)$ and diagrams of  \pak$(r)$, calculated via the
tilted-rings method. The error bars on the rotation curve
correspond the mean square deviation of points in the masked
field from the model (RMS). For the regions with  strongly
perturbed kinematics (abrupt variations of  \pak, large errors in
its estimates) or for those where the velocity field lacks points
for the confident estimation of the position angle, we assumed
\linebreak \pak\,$={\rm PA}_0$. The rotation curves for the disks
observed edge-on are shown in Fig.~\ref{fig_rc2}.
Whenever possible, we compared rotation curves with the published
data on the H\,I kinematics.

\begin{table*}
\captionstyle{normal} \caption{Galaxy parameters}
\label{tab_res}
\medskip
\begin{tabular}{l|r@{$\,\pm\,$}l|r@{$\,\pm\,$}l|c|r|c|c}\hline
\multicolumn{1}{c|}{Name} &\multicolumn{2}{c|}{$V_{\rm sys},$~km\,s$^{-1}$}  & \multicolumn{2}{c|}{${\rm PA}_0$, deg} &  $i_0$, deg & $V_{\rm max},$~km\,s$^{-1}$ & $r_{\rm max}$, arcsec &  Note  \\
\hline
CGCG\,269-049 &  $139$&$1$  &  \multicolumn{2}{c|}{$306$}      &   $43$      &   $9.8\pm0.9$ &  $45$--$60$  & H\,I \\
DDO\,53       &   $27$&$2$  &  \multicolumn{2}{c|}{$131$}      &   $27$      &  $16.4\pm2.6$ & $106$--$153$ & H\,I \\
DDO\,68       &  $511$&$1$  &  \multicolumn{2}{c|}{ $17$}      &   $65$      & $>57.2\pm2.2$ & $228$        &   H\,I \\
DDO\,99       &  $249$&$1$  &  \multicolumn{2}{c|}{ $70$}      &   $52$      &  $11.7\pm1.4$ &  $42$--$57$  &     \\
DDO\,125      &  $200$&$1$  &  \multicolumn{2}{c|}{$128$}      &   $63$      & $>17.8\pm3.0$ & $134$        &   H\,I \\
DDO\,190      &  $153$&$2$  &  \multicolumn{2}{c|}{$149$}      &   $60$      & $>24.7\pm2.5$ &  $90$        &   H\,I \\
KK\,149       &  $410$&$2$  &  $~157$&$5$                       &   $58$      &  $26.2\pm2.0$ &  $18$--$26$  &     \\
KKH\,12       &  \multicolumn{2}{c|}{--} &  \multicolumn{2}{c|}{--}      &    --       &   -- ~~~~~~~  &   --         &     \\
KKH\,34       &  \multicolumn{2}{c|}{--} &  \multicolumn{2}{c|}{--}      &    --       &   -- ~~~~~~~  &   --         &     \\
KKR\,56       &  \multicolumn{2}{c|}{--} &  \multicolumn{2}{c|}{--}      &    --       &   -- ~~~~~~~  &   --         &     \\
UGC\,231        &  $835$&$2$  &   $~56$&$3$                       &   $90$      &  $92.8\pm2.2$ & $120$--$158$ &     \\
UGC\,891        &  $637$&$1$  &   $~53$&$2$                       &   $65\pm3$  & $>60.0\pm1.6$ & $149$        &   H\,I \\
UGC\,1281       &  $140$&$1$  &  $~220$&$1$                       &   $90$      &  $56.4\pm1.4$ & $179$--$194$ &   H\,I \\
UGC\,1501       &  $183$&$2$  &    $~1$&$2$                       &   $75$      &  $47.5\pm1.5$ & $120$--$160$ &     \\
UGC\,1924       &  $586$&$1$  &  $~181$&$1$                       &   $90$      &  $50.6\pm1.8$ &  $45$--$53$  &     \\
UGC\,3476       &  $439$&$3$  &   $~60$&$3$                       &   $90\pm$   &   $8.8\pm1.9$ &  $22$--$30$  &     \\
UGC\,3672       &  $984$&$3$  &  \multicolumn{2}{c|}{$138$}      &   $56$      &  $67.8\pm1.7$ &  $51$--$53$  &      \\
UGC\,5423       &  $354$&$1$  &  $~319$&$3$                       &   $56\pm5$  &  $24.8\pm0.9$ &  $20$--$39$  &   H\,I \\
UGC\,5427       &  $484$&$3$  &  $~281$&$4$                       &   $55\pm5$  &  $54.1\pm2.8$ &  $31$--$41$  &     \\
UGC\,6456       & $-116$&$2$  &  \multicolumn{2}{c|}{$169$}      &   $66$      &  $15.0\pm0.7$ &  $67$--$82$  &   H\,I \\
UGC\,7611       &  $487$&$1$  &   $~42$&$2$                       &   $77$      &  $51.5\pm5.2$ &  $37$--$39$  &     \\
UGC\,8508       &   $60$&$2$  &  $~126$&$3$                       &   $51\pm3$  &  $32.6\pm1.2$ & $121$--$157$ &   H\,I \\
UGC\,8638       &  $275$&$1$  &  \multicolumn{2}{c|}{ $87$}      &   $49$      &  $11.3\pm2.2$ &  $30$--$34$  & \\
UGC\,11583      &  $113$&$2$  &  $~268$&$1$                       &   $90$      &  $46.7\pm1.6$ & $124$--$144$ &   H\,I \\
UGC\,11425      & $-167$&$2$  &  $~255$&$5$                       &   $35$      &  $37.1\pm1.8$ &  $21$--$28$  &     \\
UGC\,12713      &  $285$&$2$  &  \multicolumn{2}{c|}{$225$}      &   $72$      &  $14.9\pm2.9$ &  $13$--$27$  &     \\
UGCA\,92      & $-108$&$2$  &   $~55$&$5$                       &   $56$      & $>12.5\pm4.2$ &  $73$        &     \\
UGCA\,292     &  $297$&$1$  &  \multicolumn{2}{c|}{ $35$}      &   $45$      &  $21.9\pm1.7$ & $113$--$226$ &   H\,I \\
\hline
\end{tabular}
\end{table*}

Table~\ref{tab_res} lists the $V_{\rm sys}$, ${\rm
PA}_0$, and $i_0$ parameters we have adopted. The  error range is
marked only for the parameters we measured from the ionized gas
velocity fields. There the maximum rotation velocity $V_{\rm max}$
is also given. If the rotation curve achieves a clear plateau,
then $V_{\rm max}$ was found as the average value of $V_{\rm rot}$
in this range of radii ($r_{\rm max}$). However, if the growth of
rotation velocity did not stop in the outer parts of the disk,
then the velocity of the most extreme point was taken for $V_{\rm
max}$. In this case the sign  ``$>$'' shows that this is only the
lower limit, and $r_{\rm max}$ corresponds to the distance of the
considered point from the center of the galaxy.

If there was a more extended  H\,I rotation curve present, the
estimates of $V_{\rm max}$ were based on it, and in this case the
last column of the table has a  ``H\,I'' note.

\section{NOTES ON INDIVIDUAL GALAXIES}
\label{sec_notes}

\textit{CGCG\,269-049}. Emission of ionized gas is observed only
in the central part of the galaxy. The  line of sight velocity gradient
related with circular rotation is hardly noticeable. The velocity
field is dominated by peculiar motions, most likely related to
the nuclear H\,II region (outflows or an expanding shell).
Meanwhile, according to~\cite{Begum2006}, the H\,I
kinematics at large distances from the center follows regular
rotation, although the disturbance from the star-forming region
is also noticeable in the H\,I velocity field. We took the same
value of $i_0$ that was measured in~\cite{Begum2006}
from the H\,I maps, it is well consistent with the optical image
of the galaxy. The ${\rm PA}_0$ estimate is taken from the
\mbox{HyperLeda}.\footnote{http://leda.univ-lyon1.fr} The
model created with these parameters indicates that the ionized gas
rotation velocity is formally very low, less than $2$~km\,s$^{-1}$
for \mbox{$r<12''$}. We have constructed the H\,I rotation curve,
shown for comparison in Fig.~\ref{fig_rc1}, via the
tilted-rings method from the velocity field, kindly provided by
the authors of~\cite{Begum2006}. Here for
\mbox{$r<30''$} a field with the resolution of
\mbox{$\mbox{beam}=28''\times24''$} was used, while for the outer
regions they have used the data with a lower resolution
(\mbox{$\mbox{beam}=42''\times37''$}).

\textit{DDO\,53}. We have earlier discovered several expanding
shells in this galaxy~\citep{MoisLoz2012}. At the first
glance, the velocity field looks chaotic. But after masking
peculiar regions it reveals a regular line of sight velocity gradient,
consistent with the pattern observed with a lower spatial
resolution in the H\,I \citep{Begum2006,
Oh2011}. The orientation parameters were taken based on
the measurements of \cite{Oh2011} from the
neutral hydrogen velocity field. Note that the measurements from
the H$\alpha$  velocity field give a close value of the position
angle: \mbox{\pak$\,=127\pm8\degr$.} The rotation curves in H\,I
and H\,II agree within errors, except for the outer regions of the
disk in H$\alpha$, where the  H\,I rotation curve reaches plateau.

\textit{DDO\,68}.    H$\alpha$ emission is  observed only in
individual H\,II regions, but the  line of sight velocity gradient is
clearly visible in the north--south direction. The disk
orientation parameters cannot be measured based on our data. We
hence took the photometric estimates
of~\cite{Hunter2006} for ${\rm PA}_0$ and $i_0$, which
are in a satisfactory agreement with the H\,I maps given
in~\cite{Ekta2008}. The map of residual velocities
(Fig.~\ref{fig_rc1}) reveals notable peculiar velocities
only in the regions north of the center, at $r\approx10$--$40''$.
This might be caused by the imperfections of our model, since due to
the lack of significant points in this region, solid-body
rotation, followed by reaching the plateau, was assumed.
Unfortunately, the authors of~\cite{Ekta2008} give an
H\,I rotation velocity  estimate only at two points on the
opposite sides of the nucleus. A comparison with the measurements
of~\cite{Hoffman1996}, based on the maps obtained at the
Arecibo radio telescope, shows that the H\,II rotation velocity at
the same $r$ is systematically lower than the  neutral hydrogen
data \citep[][we   recomputed their date to $i_0=65\degr$]{Hoffman1996}. The difference is possibly caused by the lack
of H\,II regions located along the major axis of the galaxy.
Hence, the uncertainty in the ${\rm PA}_0$ estimate leads to a
decrease in $V_{\rm rot}$. Besides, according
to~\cite{Ekta2008}, the morphological and kinematic
features of H\,I suggest that the galaxy has experienced a recent
interaction leading to the perturbation of its gaseous disk.

\textit{DDO\,99}. The disk orientation parameters are taken from
the photometric estimates of ~\cite{Fingerhut2010}. The
velocity field is well filled, the velocities of the majority of
regions of the disk correspond to the circular rotation model. The
values of \pak\ we have measured  are within the errors consistent
with the adopted ${\rm PA}_0$ (Fig.~\ref{fig_rc1}).
Hence, even more surprising is the comparison with the published
H\,I maps. On the one hand, the HI  surface density distribution
of~\cite{Begum2008, Ott2012} agrees well with
the optical morphology, i.e., it is elongated along the ${\rm
PA}=60$--$70\degr$. On the other hand, in the H\,I velocity field
\citep[see Fig.~17 in][]{Ott2012}), \pak\ is turned almost
perpendicularly to the major axis in the density distribution so
that   \pak$\,=130$--$150\degr$. And the rotation velocity is at
least   $25$--$30$~km\,s$^{-1}$ which is significantly higher than
our estimates from H\,II. This discrepancy between the morphology
and kinematics can indicate that the disk of neutral hydrogen is
non-stationary, it was formed relatively recently as a result of
interaction or external accretion and rotates in a plane, strongly
inclined to the main galaxy. At that, the orbits of gas clouds  in
the inner regions have already precessed in the plane of the
stellar disk which is exactly what we see in the \mbox{ionized
gas.}

\textit{DDO\,125}.  The H$\alpha$ image is flocky, numerous
expanding shells are identified in it~\citep{MoisLoz2012}. At the same time, radial
velocities are in a reasonable agreement with the circular
rotation model. The galaxy has been  mapped in H\,I many times
\citep[see][ibid the references to earlier work]{Swaters2009, Begum2008, Ott2012}. We have
taken the same ${\rm PA}_0$ and $i_0$ that were measured
in~\cite{Swaters2009} based on the velocity field of
neutral hydrogen. The  H\,I and H\,II rotation curves agree within
errors. Relative growth of  rotation velocity in   H$\alpha$ by
$r>70''$ is likely an artifact related with non-circular motions
in the northwestern star-forming region.

\textit{DDO\,190}. At $r\le20''$  rotation of  ionized gas is
almost not observed, since peculiar motions dominate there. At
large distances from the center,  the velocity field follows the
circular rotation model with the orientation parameters fixed in
the  H\,I observations  of~\cite{StilIsrael2002}. The
residual velocity field reveals an area along the minor axis of
the galaxy as well as separate ionized shells. The H\,II rotation
curve reaches plateau at \mbox{$r> 30''$,} while a more extended
curve in H\,I shows further growth.

\textit{KK\,149}.   Ionized gas rotation is almost lacking in the
innermost region  \mbox{($r\le5''$)}. At larger distances, the
velocity field follows circular rotation. The residual velocity
field has noticeable disturbances associated with the shells
around the H\,II regions. We took the HyperLeda data for $i_0$
because from the kinematics the inclination  is estimated with a
great uncertainty. At the same time, the positional angle
measured from the velocity field is in a good agreement with the
available photometric estimates.

\textit{KKH\,12},   \textit{KKH\,34},  and  \textit{KKR\,56}. Only
separate compact H\,II regions are observed here. The rotation
model failed to be created.

\textit{UGC\,231}.   After subtracting the  ``rotating cylinder''
 model from the observed velocity field, the residual velocities in the
northwestern part of the disk are systematically lower than those
in the south-east. This most likely  implies  small deviations
($0.5$--$1\degr$)  of the inclination from $i = 90\degr$. The
rotation curve we constructed is in a good agreement with the
measurements based on ionized gas
from~\cite{deBlok2002}. At that, our measurements extend
farther on, to $r\approx180''$. According to the observations
of~\cite{Rhee1996}, the H\,I disk is observed up to
$r\approx300''$, where the maximum rotation velocity amounts to
$98\pm3$~km\,s$^{-1}$ which is close to our estimate.

\textit{UGC\,891}. \cite{vanZee1997} have
obtained  from the H\,I  observations \mbox{$i=60\pm5\degr$} and
\mbox{${\rm PA}=48\pm2\degr$}, what is close to the disk
orientation parameters we have measured. As in the cases of
DDO\,190 and KK\,149, rotation of ionized gas in the central
region  is almost imperceptible here. At    $r>20''$, the H\,II
kinematics follows the model of   regular rotation. The velocities
of external isolated H\,II regions are consistent with a more
extended rotation curve in H\,I~\citep{vanZee1997}.

\textit{UGC\,1281}. A good agreement with the approach of the
rotating cylinder. The H\,II  rotation curve  perfectly fits  with
the measurements in  H\,I~\citep{Swaters2009}.

\textit{UGC\,1501}.  From the analysis of the velocity field, we
have obtained $i=65\pm5\degr$,   ${\rm PA}=1\pm2\degr$. But this
$i$ value looks clearly underestimated, since the outer galaxy
isophotes correspond to a greater inclination. Therefore, we have
adopted  $i_0=75\degr$ according to the photometry
in~\cite{vanZee2000}. The ionized gas velocity field was
also presented by~\cite{Kandalyan2003}, but our
measurements are deeper. The rotation velocity step at $r>170''$
was not included in the
 $V_{\rm max}$ estimate, since it corresponds   only to one most
external   H\,II region north of the nucleus.

\textit{UGC\,1924}. A good agreement with the approach of the
rotating cylinder. The $V_{\rm max}$ estimate  is made by the
outer H\,II region, since unlike in UGC\,1501, these points
correspond to  the linear extrapolation of the rotation curve at
smaller $r$.

\textit{UGC\,3476}.  It is possible that  the inclination of the
galaxy differs by several degrees from the accepted \mbox{$i_0 =
90\degr$.} The residual velocity field  at the boundaries of the
bright H\,II regions reveals noticeable local deviations of up to
$20$--$30$~km\,s$^{-1}$. The measured maximum rotation velocity is
2--3 times lower than that expected from the Tully-Fisher relation
for a galaxy with $M_B = -14$. The rotation velocity most likely
reaches its maximum outside the \mbox{H$\alpha$}-emitting disk.

\textit{UGC\,3672}.  The disk parameters are taken according to
the photometric estimates of~\cite{vanZee2000}  because
the velocity field is strongly perturbed. Rotation with \pak\
close to ${\rm PA}_0$  is observed only for  \mbox{$r> 23''$}. The
direction of the   velocity gradient in the central region of
the galaxy   varies the way that \pak\ sharply decreases
(Fig.~\ref{fig_rc1}). Consider the possible explanations
of this kinematics:

\begin{enumerate}

\item   Significant non-circular motions of  gas. This way,
\pak\ may dramatically change  in the case of radial motions under
the effect of the bar non-axisymmetric potential \citep[see discussion and references
in][]{Moiseev2004}). But even in this case, a change
in \pak\ by almost $ 90\degr$ is   extraordinary. In addition,
the bar is not visible in the optical images. The assumption of
the outflow of gas under the effect of vigorous star formation
leaves doubts since there are no bright  H\,II regions in the
core. Furthermore, in this case we should expect a significant
increase  of gas velocity dispersion or even the splitting of
the  H$\alpha$  line profile which is not observed.

\item  Rotation in the plane different from the stellar disk of
the galaxy. Our estimate shows that the circumnuclear velocity field
can be described in the assumption that ${\rm PA} = 84\pm6\degr$.
This means that there is  an internal inclined or even polar disk.
Such structures are observed in a number of nearby galaxies,
including dwarf \citep[see][for  review]{Moiseev2012}).
However, the  brightness distribution in the  line of  H$\alpha$
for the central region of UGC\,3672 corresponds to   \mbox{${\rm
PA}=120$--$150\degr$}  which is closer to the orientation of the
major axis of the stellar disk than to $\rm PA$ of the presumed
polar structure.

\item  The observed structure is not stationary, it is a result of
the tidal effect or of the capture of the destroyed companion's
matter. It may possibly be a polar structure in the process of
formation.

\end{enumerate}

\textit{UGC\,5423}.  The orientation parameters we have obtained
coincide within the errors  with the measurements
of~\cite{Daigle2006}, based on the ionized gas velocity
field. The rotation curve we have constructed extends further than
in~\cite{Daigle2006}. \cite{Oh2011}
obtained the mean value of $i_0=44\degr$ from  the H\,I velocity
field, but they  noted that the inclination of the disk expected
from the  of Tully-Fisher baryon relationship has to be much
larger and amount to $59\degr$. Therefore, we consider  our $i_0=56\degr$
estimate to be more correct and better corresponding to the
isophote ellipticity in the image of the galaxy. The H\,I rotation
curve, demonstrated in Fig.~\ref{fig_rc1}, was
recalculated from~\cite{Oh2011} to the taken $i_0$.
Only the points up to  $r = 1$~kpc ($39''$) are shown there, as
at large distances from the center, according to the authors,
their estimates are uncertain.

\textit{UGC\,5427}.  A good agreement with the approach of
circular rotation, except for certain regions of low surface
brightness. A decrease of rotation velocity at $r> 40''$ is most likely an
artifact, since there exist measurements only for the eastern half of the
disk.

\textit{UGC\,6456}.  A well-known blue compact galaxy with
numerous ionized shells around the star forming regions
\citep[][ibid the references to earlier work]{Simpson2011, MoisLoz2012}. ${\rm PA}_0$, $i_0$ were taken
according to the photometric estimate of~\cite{Simpson2011}. The same article gives the H\,I
velocity field. However, the orientation parameters based on the
kinematics of neutral hydrogen are determined with great
uncertainty, since the velocity field is perturbed and
asymmetrical which may be the result of tidal interaction.
However, the maximum rotation velocity from H\,I and H\,II prove
to be close (see Fig.~\ref{fig_rc1}, where the H\,I
rotation curve for the north-west half of the galactic disk is
shown for comparison). The deviations of  line of sight velocities in
H$\alpha$ from the circular rotation model are related both with
the individual expanding shells and with the overall asymmetry of
the gaseous disk, noted in~\cite{Simpson2011}.

\textit{UGC\,7611}.  \cite{Kaisin2008}  have discovered that the
emission of ionized gas  is observed not only around the regions
of star formation but also far beyond the strongly inclined disk
of the galaxy. Further spectroscopic
observations~\citep{Moiseev2010_4460} have shown that the
extended emission nebula is composed of galactic wind, i.e., gas
ejected from the plane of the disk as a result of a cumulative
effect of stellar winds and supernova explosions. At that, in the
disk itself, according our  velocity, the kinematics of ionized gas is in a good agreement with the
circular rotation model. The $i_0$ value is taken according to the
photometric estimate from~\cite{Moiseev2010_4460}, ${\rm
PA}_0$ is calculated from the velocity field. In the regions where
the galactic wind is present,   line of sight velocities significantly (up
to $50$~km\,s$^{-1}$) differ from what the
extrapolation of the circular rotation model. Obviously, models
that are kinematically more complex need to be applied here,
taking into account the geometry and speed of the outflow.

\textit{UGC\,8508}.  If the pronounced velocity gradient observed
in the H\,II velocity field results from circular rotation, it
corresponds to \mbox{\pak$\,=80$--$100\degr$}. At that, the image
of the galaxy itself in the line of H$\alpha$ is extended along
this direction. However, such a $\rm PA$ drastically differs from
the outer isophote major axis orientation in the optical image of
the galaxy \mbox{(${\rm PA} = 123\degr$, HyperLeda).} The density
distribution  in H\,I at large distances from the center is also
extended along ${\rm PA}=115$--$120\degr$ \citep[see the maps
in][]{Begum2008, Ott2012}. Thus, the
situation here is similar to the above-described cases of the
disagreement of kinematic parameters in the inner and outer
regions of DDO\,99 and UGC\,3672. Compared with these galaxies,
UGC\,8508 reveals the most regular motions of ionized gas. We have
earlier identified  two giant ionized shells as well as a nebula
around a candidate LBV star~\citep{MoisLoz2012}. But
their contribution to the perturbation of the velocity field   is
small, and while constructing the model, these regions were
excluded from consideration. We used  the tilted-rings method to
make an analysis of the H\,I velocity field obtained at the VLA
with the resolution of $14\times12''$. The orientation parameters
for the outer regions of the disk ($r>80''$) are given in
Table~\ref{tab_res} and agree well with the above
photometric estimates. However, at $r<50''$ there is a sharp
reversal of \pak, while its measurements in H\,I and H\,II
coincide. We have therefore adopted for UGC\,8508 a model of a gas
layer warped in the inner part, ensuing from the plane of the
stellar disk of the galaxy. Similar strong internal warps are
observed, for example, in the gaseous disks of the lenticular
galaxy NGC\,2685~\citep{Jozsa2009} and the dwarf galaxy
Mrk\,370~\citep{Moiseev2011EAS}. If we accept for the
disk of ionized gas of UGC\,8508  ${\rm PA}\approx90\degr$,
$i\approx50\degr$, then the calculation of the angle between the
planes of the stellar disk and the H\,II disk \citep[see the formula
in][]{Moiseev2008}  would yield two values: $28\degr$
and $86\degr$. The latter value corresponds almost exactly to the
orthogonal mutual orientation, i.e., UGC\,8508 possibly includes
an internal polar disk.

\textit{UGC\,8638}.  The velocity field is  well filled by the
H$\alpha$ emission, but it is quite difficult to separate the
regular rotation and peculiar motions. The orientation parameters
of the disk are taken from the HyperLeda.

\textit{UGC\,11425}.  The velocity field is dominated by regular
circular rotation.  The inclination  is adopted in accordance
with  the  photometry from~\cite{Hunter2006}, ${\rm PA}_0$
was refined from the velocity field.

\textit{UGC\,11583}. The velocity field is  described well by the
 rotating cylinder model, but the  H\,II rotation curve does not
always coincide  with a more extended  H\,I rotation curve
from~\cite{BegumChengalur2004}. Firstly, the plateau in
H\,II  is reached at the smaller distances    $r=40''$, which is
possibly related with a better spatial resolution of our data. A
faster rotation velocity in H$\alpha$ at \mbox{$r=60$--$100''$}
corresponds to the chain of   H\,II regions at the western edge of
the disk. It is difficult to understand why their radial
velocities are by $15$--$20$~km\,s$^{-1}$  greater than the
maximum rotation velocity. They   possibly belong to a tidal
structure.

\textit{UGC\,12713}.   The disk orientation parameters are taken
according to the analysis of  H\,I maps
in~\cite{Noordermeer2007}. The ionized gas velocity
field reveals a notable reversal of isovels. The velocities on the
southern edge of the disk vary from the circular model by more
than $30$~km\,s$^{-1}$. The  tidal perturbation is the most likely
explanation for such peculiar motions.

\textit{UGCA\,92}.   The H$\alpha$ emission is mainly observed  in
numerous expanding ionized shells~\citep{MoisLoz2012}.
The contribution of  regular rotation in the velocity field is
negligible, but we able to calculate the rotation curve. The   $i_0$ is taken in
accordance with the observations in
H\,I~\citep{Begum2008}. The  H\,I density distribution
outside the optical disk is complex,   external H\,I isodenses are
turned almost perpendicularly to the optical image. For   ${\rm
PA}_0$  our estimate is given from the isophotes of the POSS\,2
image.

\textit{UGCA\,292} The galaxy with an extended disk of neutral
hydrogen, the corresponding maps are shown
in~\citet{Lo1993,Young2003}. Unfortunately, the
H$\alpha$ line reveals  only a few  H\,II regions near the center
of the galaxy. Rotation velocity of  ionized gas does not exceed
$10$~km\,s$^{-1}$  which is two times smaller than the maximum
velocity of rotation in H\,I, shown in
Fig.~\ref{fig_rc1} according
to~\cite{Hoffman1996} (the data are recomputed to the
taken  $i_0$). The $i_0$ estimate is adopted
from~\cite{Young2003}, and ${\rm
PA}_0$---from~\cite{Lo1993}.

\section{CONCLUSION}
\label{sec_conclusion}

We have constructed and analyzed  the ionized gas velocity fields
in 28 nearby dwarf galaxies. These results (rotation curves,
maximum rotation velocity estimates) can be useful for studying
the mass distribution in these objects. Specifically, they are
already  used in our paper~\citep{MoiseevTikhonov2014}.
Note some points concerning the internal kinematics of galaxies
considered:
\begin{enumerate}

\item  The ionized gas velocity perturbation amplitude in dwarf
galaxies  under the effect of the  star formation   may even
exceed the rotation velocity. However, in the majority of cases
(in 25 galaxies) we can successfully identify the kinematic
component associated with regular circular rotation.  The  H\,II
rotation curves we have constructed are generally in a good
agreement with  data measured from neutral hydrogen, taking into
account  the spatial resolution difference.

\item  In several cases,  rotation pattern of  ionized gas is
very different from the one that should be expected from the
orientation of external optical  isophotes or the large-scale
 H\,I velocity field.  In DDO\,99, for instance, the inner part of the gaseous disk,
observed in the   H$\alpha$ emission line, lies in the stellar disk plane,
while the external gas, observed in the 21 cm line is out of this
plane. These structures are most likely the result of interaction
with a companion or external accretion  of gas. The traces of
tidal perturbations  are visible in the velocity fields of
UGC\,6456, UGC\,11583, UGC\,12713, and UGC\,3672. In the latter
case, it seems that the inner part of the disk of ionized gas
rotates at a large angle to the stellar disk. The variation of
orbit orientation of gas clouds can be most thoroughly traced in
the UGC\,8508. Here the external  H\,I disk coincides with the
plane of the stellar disk, but in the inner region we observe a
warp of the gas layer, traced both in the H\,I and  H\,II. It is
possible that the entire ionized gas in the UGC\,8508 rotates in
the plane orthogonal to the stellar disk of the galaxy.
\end{enumerate}

\begin{acknowledgements}
This work was carried out with the financial support of the
non-profit Dynasty Foundation, the program of the Ministry of
Education and~Science of Russian Federation (project 8523), the
RFBR grant no.~13-02-00416, and the Basic Research Program of the
RAS Division of Physical Sciences OFN-17 ``Active Processes in the
Galactic and Extragalactic Objects. We used the NASA/IPAC (NED) database of extragalactic data,
managed by the Jet Propulsion Laboratory of the California
Institute of Technology under the contract with the NASA (USA),
and the HyperLeda database. The paper is based on the
observational data of the 6-m telescope of the   Special
Astrophysical Observatory of RAS operating with the financial support
from the RF Ministry of Education and Science (state contracts
no.~16.518.11.7073 and 14.518.11.7070). The author thanks Jayaram
Chengalur and Ayesha Begum, who have kindly presented the
original data published in their papers, and Anatoly Klypin and
Simon Pustilnik for useful discussions.
\end{acknowledgements}



\begin{thebibliography}{99}
\bibitem[\protect\citeauthoryear{{Afanasiev} \& {Moiseev}}{2005}]{AfanasievMoiseev2005}
{Afanasiev}~V.L., \& {Moiseev}~A.V., 2005, Astronomy Letters, {31}, 194

\bibitem[\protect\citeauthoryear{{Begum} \& {Chengalur}}{2004}]{BegumChengalur2004}
{Begum}~A., \& {Chengalur}~J.N., 2004, \aap, {424}, 509

\bibitem[\protect\citeauthoryear{{Begum} et al.}{2006}]{Begum2006}
{Begum}~A., {Chengalur}~J.N., {Karachentsev}~I.D., et al., 2006, \mnras, {365}, 1220

\bibitem[\protect\citeauthoryear{{Begum} et al.}{2008}]{Begum2008}
{Begum}~A., {Chengalur}~J.N., {Karachentsev}~I.D., et al., 2008, \mnras, {386}, 1667

\bibitem[\protect\citeauthoryear{{Daigle} et al.}{2006}]{Daigle2006}
{Daigle}~O., {Carignan}~C., {Amram}~P., et al., 2006, \mnras, {367}, 469

\bibitem[\protect\citeauthoryear{{de~Blok} \& {Bosma}}{2002}]{deBlok2002}
{de~Blok}~W.J.G., \& {Bosma}~A., 2002, \aap, {385}, 816


\bibitem[\protect\citeauthoryear{{Doroshkevich}, {Lukash} \& {Mikheeva}}{{Doroshkevich} et al.}{2012}]{Doroshkevich2012UFN}
{Doroshkevich}~A.G., {Lukash}~V.N., \& {Mikheeva}~E.V., 2012, Physics Uspekhi, {55}, 3

\bibitem[\protect\citeauthoryear{{Ekta}, {Chengalur} \& {Pustilnik}}{{Ekta} et al.}{2008}]{Ekta2008}
{Ekta}~B., {Chengalur}~J.N., \&  {Pustilnik}~S.A., 2008, \mnras, {391}, 881

\bibitem[\protect\citeauthoryear{{Epinat}, {Amram} \& {Marcelin}}{{Epinat} et al.}{2008}]{Epinat2008}
{Epinat}~B., {Amram}~P., \& {Marcelin}~M., 2008, \mnras, {390}, 466

\bibitem[\protect\citeauthoryear{{Fingerhut} et al.}{2010}]{Fingerhut2010}
{Fingerhut}~R.L., {McCall}~M.L., {Argote}~M., et al., 2010, \apj, {716}, 792

\bibitem[\protect\citeauthoryear{{Hoffman} et al.}{1996}]{Hoffman1996}
{Hoffman}~G.L., {Salpeter}~E.E., {Farhat}~B., et al., 1996, \apjs, {105}, 269

\bibitem[\protect\citeauthoryear{{Hunter} \& {Elmegreen}}{2006}]{Hunter2006}
{Hunter}~D.A., \&  {Elmegreen}~B.G., 2006, \apjs, {162}, 49

\bibitem[\protect\citeauthoryear{{Hunter} et al.}{2012}]{Hunter2012}
{Hunter}~D.A., {Ficut-Vicas}~D., {Ashley}~T., et al., 2012, \aj, {144}, 134

\bibitem[\protect\citeauthoryear{{Jєzsa} et al.}{2009}]{Jozsa2009}
{Jєzsa}~G.I.G., {Oosterloo}~T.A., {Morganti}~R., et al., 2009, \aap, {494}, 489

\bibitem[\protect\citeauthoryear{{Kaisin} \& {Karachentsev}}{2008}]{Kaisin2008}
{Kaisin}~S.S., \& {Karachentsev}~I.D., 2008, \aap, {479}, 603

\bibitem[\protect\citeauthoryear{{Kamphuis} et al.}{2007}]{Kamphuis2007}
{Kamphuis}~P., {Peletier}~R.F., {Dettmar}~R.-J., et al., 2007, \aap, {468}, 951

\bibitem[\protect\citeauthoryear{{Kandalyan}, {Khassawneh} \& {Kalloghlian}}{{Kandalyan} et al.}{2003}]{Kandalyan2003}
{Kandalyan}~R.A., {Khassawneh}~A.M., \&  {Kalloghlian}~A.T., 2003, Astrophysics, {46}, 74

\bibitem[\protect\citeauthoryear{{Lozinskaya} et al.}{2006}]{Lozinsk2006}
{Lozinskaya}~T.A., {Moiseev}~A.V., {Avdeev}~V.Y., {Egorov}~O.V., 2006, Astronomy Letters, {32}, 361

\bibitem[\protect\citeauthoryear{{Lo}, {Sargent} \& {Young}}{{Lo} et al.}{1993}]{Lo1993}
{Lo}~K.Y., {Sargent}~W.L.W., \& {Young}~K., 1993, \aj, {106}, 507

\bibitem[\protect\citeauthoryear{{Moiseev}}{2002}]{Moiseev2002ifp}
{Moiseev}~A.V., 2002, BSAO, {54}, 74

\bibitem[\protect\citeauthoryear{{Moiseev}}{2008}]{Moiseev2008}
{Moiseev}~A.V., 2008, Astrophysical Bulletin, {63}, 201

\bibitem[\protect\citeauthoryear{ Moiseev   }{2011}]{Moiseev2011EAS}
{Moiseev}~A., 2011, in \emph{EAS Publications Series}, Vol.~48, 115

\bibitem[\protect\citeauthoryear{{Moiseev}}{2012}]{Moiseev2012}
{Moiseev}~A.V., 2012, Astrophysical Bulletin, {67}, 147

\bibitem[\protect\citeauthoryear{{Moiseev} \& {Egorov}}{2008}]{MoiseevEgorov2008}
{Moiseev}~A.V., \&  {Egorov}~O.V., 2008, Astrophysical Bulletin, {63}, 181

\bibitem[\protect\citeauthoryear{{Moiseev} \& {Lozinskaya}}{2012}]{MoisLoz2012}
{Moiseev}~A.V., \& {Lozinskaya}~T.A., 2012, \mnras, {423}, 1831

\bibitem[\protect\citeauthoryear{{Moiseev}, {Vald\'{e}s} \& {Chavushyan}}{{Moiseev} et al.}{2004}]{Moiseev2004}
{Moiseev}~A.V., {Vald\'{e}s}~J.R., \& {Chavushyan}~V.H., 2004, \aap, {421}, 433

\bibitem[\protect\citeauthoryear{{Moiseev}, {Karachentsev} \& {Kaisin}}{{Moiseev} et al.}{2010a}]{Moiseev2010_4460}
{Moiseev}~A., {Karachentsev}~I., \& {Kaisin}~S., 2010, \mnras, {403}, 1849

\bibitem[\protect\citeauthoryear{{Moiseev}, {Pustilnik} \& {Kniazev}}{{Moiseev} et al.}{2010b}]{Moiseev2010}
{Moiseev}~A.V., {Pustilnik}~S.A., \& {Kniazev}~A.Y., 2010, \mnras, {405}, 2453

\bibitem[\protect\citeauthoryear{{Moiseev}, {Tikhonov} \& Klypin}{2014}]{MoiseevTikhonov2014}
{Moiseev}~A.V., {Tikhonov}~A.V., \&  Klypin A., 2014, \mnras, submitted


\bibitem[\protect\citeauthoryear{Noordermeer \& van~der~Hulst}{2007}]{Noordermeer2007}
{Noordermeer}~E., \&  {van~der~Hulst}~J.~M., 2007, \mnras, {376}, 1480

\bibitem[\protect\citeauthoryear{{Oh} et al.}{2011}]{Oh2011}
{Oh}~S.-H., {de Blok}~W.J.G., {Brinks}~E., et al., 2011, \aj, {141}, 193

\bibitem[\protect\citeauthoryear{{Ostlin} et al.}{1999}]{Ostlin1999}
{Ostlin}~G., {Amram}~P., {Masegosa}~J., et al., 1999, \aaps, {137}, 419

\bibitem[\protect\citeauthoryear{{Ott} et al.}{2012}]{Ott2012}
{Ott}~J., {Stilp}~A.M., {Warren}~S.R., et al., 2012, \aj, {144}, 123

\bibitem[\protect\citeauthoryear{{Rhee} \& {van~Albada}}{1996}]{Rhee1996}
{Rhee}~M.-H., \& {van~Albada}~T.S., 1996, \aaps, {115}, 407

\bibitem[\protect\citeauthoryear{{Simpson} et al.}{2011}]{Simpson2011}
{Simpson}~C.E., {Hunter}~D.A., {Nordgren}~T.E., et al., 2011, \aj, {142}, 82

\bibitem[\protect\citeauthoryear{{Staveley-Smith}, {Davies} \& {Kinman}}{{Staveley-Smith} et al.}{1992}]{Staveley-Smith1992}
{Staveley-Smith}~L., {Davies}~R.D., \& {Kinman}~T.D., 1992, \mnras, {258}, 334

\bibitem[\protect\citeauthoryear{{Stil} \& {Israel}}{2002}]{StilIsrael2002}
{Stil}~J.M., \& {Israel}~F.P., 2002, \aap, {389}, 42

\bibitem[\protect\citeauthoryear{{Swaters} et al.}{2009}]{Swaters2009}
{Swaters}~R.A., {Sancisi}~R., {van~Albada}~T.S., {van~der~Hulst}~J.M., 2009, \aap, {493}, 871

\bibitem[\protect\citeauthoryear{{van~Zee} et al.}{1997}]{vanZee1997}
{van~Zee}~L., {Maddalena}~R.J., {Haynes}~M.P., et al., 1997, \aj, {113}, 1638

\bibitem[\protect\citeauthoryear{van~Zee}{2000}]{vanZee2000}
 van~Zee~L., 2000, \aj, {119}, 2757


\bibitem[\protect\citeauthoryear{{Young} et al.}{2003}]{Young2003}
{Young}~L.M., van~Zee~L., {Lo}~K.Y., et al., 2003, \apj, {592}, 111

\bibitem[\protect\citeauthoryear{{Zasov} \& {Khoperskov}}{2003}]{ZasovKhoperskov2003}
{Zasov}~A.V., {Khoperskov}~A.V., 2003, Astronomy Letters, {29}, 437



\end{thebibliography}
\end{document}